\documentclass{svjour3}                     
\smartqed  
\usepackage{graphicx}
 \usepackage{mathptmx}      
\newcommand{\bk}{{\bf k}}
\newcommand{\bn}{{\bf n}}
\newcommand{\bp}{{\bf p}}

\newcommand{\bq}{{\bf q}}
\newcommand{\br}{{\bf r}}

\newcommand{\be}{\begin{equation}}
\newcommand{\en}{\end{equation}}
\newcommand{\eeq}{\end{equation}}
\newcommand{\bear}{\be\begin{array}}
\newcommand{\bea}{\begin{eqnarray}}
\newcommand{\eea}{\end{eqnarray}}

\def\lsim{\mathrel{\rlap{\lower4pt\hbox{\hskip1pt$\sim$}}
    \raise1pt\hbox{$<$}}}         

\def\gsim{\mathrel{\rlap{\lower4pt\hbox{\hskip1pt$\sim$}}
    \raise1pt\hbox{$>$}}}         

\begin{document}

\title{High-energy physics with particles carrying non-zero orbital angular momentum\thanks{Presented at the workshop "30 years of strong interactions", Spa, Belgium, 6-8 April 2011.}}


\author{Igor P. Ivanov
}


\institute{I. P. Ivanov\at
              IFPA, Universit\'{e} de Li\`{e}ge, All\'{e}e du 6 Ao\^{u}t 17, b\^{a}timent B5a, 4000 Li\`{e}ge, Belgium\\ and
Sobolev Institute of Mathematics, Koptyug avenue 4, 630090, Novosibirsk, Russia\\
              \email{Igor.Ivanov@ulg.ac.be}           
}

\date{Received: date / Accepted: date}

\maketitle

\begin{abstract}
Thanks to progress in optics in the past two decades, it is possible to create photons 
carrying well-defined non-zero orbital angular momentum (OAM). 
Boosting these photons into high-energy range preserving their OAM seems feasible. 
Intermediate energy electrons with OAM have also been produced recently.
One can, therefore, view OAM as a new degree of freedom in high-energy collisions
and ask what novel insights into particles' structure and interactions it can bring.
Here we discuss generic features of scattering processes involving particles with OAM in the initial state. 
We show that they make it possible to perform a Fourier analysis of a plane wave cross section 
with respect to the azimuthal angles of the initial particles, and to probe the autocorrelation function of the amplitude, 
a quantity inaccessible in plane wave collisions.

\end{abstract}

\section{Introduction}\label{section-intro}
Laser beams carrying non-zero orbital angular momentum (OAM) are well-known and routinely used in optics, \cite{OAMreview}.
The lightfield in such beams is described via non-plane wave solutions of the Maxwell equations, for example by Bessel 
or Laguerre-Gaussian beams. 
Each photon in this lightfield, which we call a {\em twisted photon}, 
carries a well-defined energy and longitudinal momentum directed along an arbitrarily chosen axis $z$ 
as well as a definite OAM projection onto this axis quantized in units of $\hbar$.
The wavefronts of such a lightfield are not planes but helices.

So far, experiments with twisted light were confined mostly to the
optical energy range. However it was recently noted that one can use Compton
backscattering of twisted optical photons off an ultra-relativistic electron beam 
to generate high-energy photons carrying non-zero OAM \cite{serbo1,serbo2}. 
Technology necessary for such convertion already exists.
In addition, successful creation of twisted electrons has also been reported recently, 
\cite{twisted-electron}. Such electrons carried the energy as
high as 300 keV and the orbital quantum number up to $m\sim 100$.
One can imagine that future progress in this field will lead to
creation of even more energetic twisted electrons and other
particles, which can then be used in scattering experiments. 
Thus, OAM can be viewed as a new degree of freedom which one can exploit in preparing
initial states of a high-energy process.
It is therefore timely to ask how such a collision can be described and what new insights 
into the properties of particles and their interactions it can bring.

In this contribution we consider generic features of scattering processes involving twisted particles
in the initial state. Specifically, we consider two cases --- single-twisted scattering (collision of a twisted state with a plane wave)
and double-twisted scattering (collision of two coaxial twisted states) --- and derive the cross sections 
of these processes in terms of the corresponding plane wave cross sections.
More details about the results presented here can be found in \cite{ivanov2011}. 

\section{Describing twisted states}
\label{section-describing}

\subsection{Scalar case}
\label{subsection-scalar}

Here we briefly summarize the formalism of Bessel-beam twisted
states introduced in \cite{serbo1} starting with the scalar field.

We first fix the $z$ axis and solve the free wave equation in
cylindric coordinates $r, \varphi_r, z$. A solution
$|\kappa,m\rangle$ with definite frequency $\omega$,
longitudinal momentum $k_z$, modulus of the transverse momentum
$|\bk|=\kappa$ and a definite $z$-projection of the orbital angular momentum $m$ has
the form
 \be
|\kappa, m\rangle = e^{-i\omega t + i k_z z} \cdot
\psi_{\kappa m}(\br)\,, \quad \psi_{\kappa m}(\br) = {e^{i m
\varphi_r} \over\sqrt{2\pi}}\sqrt{\kappa}J_{m}(\kappa r)\,,
 \label{twisted-coordinate}
  \en
where $J_m(x)$ is the Bessel function. The transverse spatial
distribution is normalized according to
 \be
\int d^2\br\, \psi^*_{\kappa' m'}(\br)\psi_{\kappa m}(\br) =
\delta_{m m'} \sqrt{\kappa\kappa'}\int rdr J_{m}(\kappa
r) J_{m}(\kappa' r) = \delta_{m m'}\,
\delta(\kappa-\kappa')\,.
 \en
A twisted state can be represented as a superposition of plane
waves:
 \be
|\kappa,m\rangle = e^{-i\omega t + i k_z z} \int {d^2\bk
\over(2\pi)^2}a_{\kappa m}(\bk) e^{i\bk \br}\,,
 \label{twisted-def}
  \en
where
 \be
a_{\kappa m}(\bk)= (-i)^m
e^{im\varphi_k}\sqrt{2\pi}\;{\delta(|\bk|-\kappa)\over
\sqrt{\kappa}}\,.\label{a-def}
  \en
This expansion can be inverted:
 \be
e^{i\bk\br} =  \sqrt{2\pi \over\kappa}
\sum_{m=-\infty}^{+\infty} i^m e^{-im\varphi_k}
|\kappa,m\rangle \,,\quad \kappa = |\bk|\,. \label{PW}
 \en
From (\ref{twisted-def}) one can simply state that the twisted state is a steady interference pattern 
arising from all plane waves
with fixed $|\bk|$ and arriving from different directions.

More details about properties of twisted states, their
normalization and phase space density can be found in
\cite{serbo2,ivanov2011}. Here we just note that although the wave
oscillation amplitude decreases at large radii, the pure Bessel twisted state
of finite amplitude is still not localized and not normalizable in the transverse plane.
Therefore, the intermediate calculations with these Bessel-beam states must be
carried out inside a large but finite cylindric volume of radius
$R$.

\subsection{Twisted photons}
\label{subsection-photons}

Description of photons carrying non-zero OAM is subtler than for scalar particles 
due to their polarization degree of freedom.
A plane wave photon with helicity $\lambda = \pm 1$ is described, in addition to the fixed four-momentum $k^\mu$, 
by an appropriately defined polarization vector $e^\mu_{\lambda}(k)$, 
with the properties $e_{\lambda\mu} k^\mu = 0$ and $e^*_{\lambda \mu} e^{\mu}_{\lambda'} = - \delta_{\lambda\lambda'}$.
In the plane wave case, the polarization vector appears as an overall factor in front of the space-time wave function:
the components of the polarization vector, which can be selected to be only transverse,
remain constant across the transverse plane orthogonal to the Poynting vector. 
The same is valid for the Stokes parameters for a general elliptic polarization state.

In the twisted case both the polarization vector of a pure helicity state and the Stokes parameters of an elliptically polarized
state acquire non-trivial spatial dependence. Even worse, the polarization vectors taken at different points cannot lie
in the same plane because the directions of the Poynting vector calculated at distinct spatial points are different.

One can represent a pure helicity twisted photon state in the coordinate space similarly to (\ref{twisted-def}):
\be
A^{\mu}_{\lambda\, \kappa m}(x) =  \sqrt{4\pi} \int {d^2\bk \over(2\pi)^2} 
\, e^\mu_{\lambda}(k)\, a_{\kappa m}(\bk) e^{-i k_\mu r^\mu}\,,
 \label{twisted-photon-def}
\en
Even with the four-potential (\ref{twisted-photon-def}) depending non-trivially on the coordinates, the gauge invariance 
in its usual definition as an invariance under $A_\mu(x) \to A_\mu(x) + \partial_\mu f(x)$ still holds.
Note however that the definition (\ref{twisted-photon-def}) should be accompanied with a prescription
of how vectors $e^\mu_{\lambda}(k)$ for different $k$ are related to each other.
Recall that for the plane waves with $\vec k = (0,0,k)$, the polarization vector is defined up to an overall phase:
\be
\vec e_{\lambda} = -{1 \over\sqrt{2}}(\lambda, i, 0) \cdot e^{i\alpha}\,,\label{e1}
\en
but the (arbitrary) $\alpha$ disappears in the matrix elements squared.
This phase shift is equivalent to the shift of the zero moment of time.

This can be repeated for each plane wave inside a twisted state. 
If the three-momentum $\vec k = \omega (\sin\theta \cos\phi,\,\sin\theta\sin\phi,\,\cos\theta)$,
we can introduce, following \cite{bliokh-light}, the unit vectors $\vec e_\theta$, $\vec e_\phi$ and construct circular polarizations as
\be
\vec e_{\lambda}(\vec k) = -{1 \over\sqrt{2}}(\lambda \vec e_\theta + i \vec e_\phi) \cdot e^{i\alpha(\phi)}\,,\label{e2}
\en
where $\alpha(\phi)$ is, in principle, an arbitrary periodic function, which, however,
does not affect the physical observables.
We choose $\alpha(\phi) = \lambda \phi$ which yields the correct paraxial limit: when $\kappa \to 0$, 
the polarization vector (\ref{e2}) turns into (\ref{e1}).

Let us also note in passing that the polarization states of twisted photons are much richer than for the plane waves.
The Stokes parameters describing the polarization locally can change from point to point, so that one must deal with
{\em polarization field} rather than polarization parameters. The polarization field is even allowed to have singularities 
at the transverse origin, where the intensity of the lightfield is zero for any $m \not = 0$.
Such singular lightfields are also well-known in optics, see e.g. \cite{singular-optics}.

At this point it is necessary to address the issue of spin-OAM separation, which (in the non-abelian case of QCD) 
is a hot topic in the high-energy physics community, especially in the context of the spin proton puzzle \cite{proton-spin}.
Here we talk about photons with a definite value of OAM and a definite helicity $\lambda$.
However this is not the spin/OAM separation that one usually has in mind due to two reasons:
(1) this is a separation of spin and OAM degrees of freedom not at the level of operators, but at the level of their average values
over certain states, and (2) the average vales of only $z$ components of these operators are involved.
At this level, the possibility to separate these two degrees of freedom is not unexpected, see e.g. a detailed discussion in \cite{leader}.
Let us also mention that for the paraxial twisted light beams the separation of OAM and helicity
is also easily derived, see \cite{spinOAMseparation}. For non-paraxial beams this issue is more tricky;
in this case the evolution of OAM and helicity in the course of beam propagation can be cast into
the form of an effective spin-orbital interaction, \cite{bliokh-light}.

\section{Single-twisted scattering}

Let us now consider a generic collision of a twisted particle with a plane wave:
\be
|\kappa,m\rangle + |PW(\bp) \rangle \to X\,.\label{single-twisted}
\en
The final system $X$ is assumed to be describable by plane waves.
The passage from the plane wave to the twisted state is given by (\ref{twisted-def})
and is applied at the level of scattering matrix: 
\be
S_{tw} = \int {d^2 \bk \over (2\pi)^2} a_{\kappa m}(\bk) S_{PW}(\bk,\bp)\,.
\en
Its square is
\bea
|S_{tw}|^2 &=& \int {d^2 \bk \over (2\pi)^2} {d^2 \bk' \over (2\pi)^2} a_{\kappa m}(\bk) a^*_{\kappa m}(\bk') 
S_{PW}(\bk,\bp)S^*_{PW}(\bk',\bp)\nonumber\\ 
&& \hspace{-1cm}\propto \int {d^2 \bk \over (2\pi)^2} {d^2 \bk' \over (2\pi)^2} a_{\kappa m}(\bk) a^*_{\kappa m}(\bk') 
\delta^{(2)}(\bk+\bp-\bp_X) \delta^{(2)}(\bk'+\bp-\bp_X) {\cal M}(\bk,\bp) {\cal M}^*(\bk',\bp)\nonumber\\ 
&& =\int {d^2 \bk \over (2\pi)^4} a_{\kappa m}(\bk) a^*_{\kappa m}(\bk) 
\delta^{(2)}(\bk+\bp-\bp_X) |{\cal M}(\bk,\bp)|^2\,.
\nonumber
\eea
Here $\bp_X$ is the transverse momentum of the final system $X$.
The last line here contains the expression that enters the plane wave cross section of the same process.
Skipping details which can be found in \cite{ivanov2011}, we give the final result which links the single-twisted 
cross section to the plane wave cross section:
\be
d\sigma_{tw} = \int {d\phi_k \over 2\pi}\, d\sigma_{PW}(\bk)\cdot {j_{PW}(\bk) \over j_{tw}}\,.\label{single-twisted-xsection}
\en
Here $j_{PW}$ and $j_{tw}$ are the plane-wave and the twisted flux functions. Discussions on subtle aspects of the definitions of the
(averaged) cross section and the flux functions for the twisted scattering can be found in \cite{serbo2,ivanov2011}.
Here we just note that the ratio of the fluxes in (\ref{single-twisted-xsection}) is very close to unity for $\kappa$ values
achievable with today's technology.

The expession (\ref{single-twisted-xsection}) is remarkable in several aspects. 
First, it is an unusual quantity in the sense that it involves averaging over the initial particle's azimuthal angle at fixed final momenta.
Second, the single-twisted cross section is $m$-independent, which proves that twisted particles scatter as easily  as plane waves.
Third, the cross section (\ref{single-twisted-xsection}) stays finite in the small $\kappa$ limit. 
Fourth, $d\sigma_{tw}$ is represented as an incoherent sum of $d\sigma_{PW}(\bk)$ for different angles $\phi_k$, 
although the initial twisted state itself is a coherent superposition. The initial coherence is lost not during the interaction
itself, but as a result of the usual condition that final states with distinct momenta do not interfere in the incoherent detectors we have.

Let us now consider the same process (\ref{single-twisted}) but assume that the twisted particle 
is in a superposition of states with different $m$, for example, $a|\kappa,m\rangle + a'|\kappa,m'\rangle$,
with $|a|^2+|a'|^2=1$.  
The calculation can be repeated yielding
\be
d\sigma = d\sigma_{tw} + 2|aa'| d\sigma^{\Delta m}_{tw}\,.
\en
where $d\sigma_{tw}$ is given by (\ref{single-twisted-xsection}) and the new term is
\be
d\sigma^{\Delta m}_{tw} = 
\int {d\phi_k \over 2\pi}\, \cos(\Delta m\,\phi_k + \alpha)\,d\sigma_{PW}(\bk)\cdot {j_{PW}(\bk) \over j_{tw}}\,,\label{single-twisted-xsection2}
\en
with $\Delta m = m-m'$ and $\alpha$ being the relative phase between $a$ and $a'$.

Looking at (\ref{single-twisted-xsection}) and (\ref{single-twisted-xsection2}) can observes that 
with a well controlled $m$ distribution of the initial twisted state one can perform the Fourier-analyzer of the plane wave cross section 
with respect to the initial azimuthal angle $\phi_k$.

What can be a potential application of this new tool? Imagine a typical elastic scattering of a probe particle on a (polarized) target of an unknown structure.
Let the initial and final transverse momenta of the probe particle be $\bk$ and $\bk'$, respectively, 
with $\bq=\bk-\bk'$ being the momentum transfer. If we are interested in the elastic formfactor, we study the $\bq^2$-dependence of the cross section
by measuring scattering at different final polar angles of the vector $\bk'$. We can be also interested in azimuthal dependence of 
the $\bk'$-distribution with respect, for example, to the transverse polarization direction $\bn$ of the target, which would allow us to measure
the $\bq\bn$ dependence.

With the new tool, one can, in principle, access the same $\bq^2$-dependent and $\bq\bn$-dependent terms with {\em fixed} final momentum $\bk'$.
Of course, the resolving power depends on how exactly the initial twisted state can be prepared,
but being a completely different experimental set-up, it might become an interesting complementary tool.

\section{Double-twisted cross section}

Let us now consider a collision of two twisted particles:
\be
|\kappa,m\rangle + |\eta,n\rangle \to X\,.
\en
Now, the scattering matrix is 
\be
S_{2tw} = \int {d^2 \bk \over (2\pi)^2}\, {d^2 \bp \over (2\pi)^2} a_{\kappa m}(\bk) 
a_{\eta n}(\bp) S_{PW}(\bk,\bp)\,,
\en
and its square contains
\bea
&&\int {d^2 \bk\,d^2 \bp\, d^2 \bk'\,d^2 \bp' \over (2\pi)^8} a_{\kappa m}(\bk) a_{\eta n}(\bp)
a^*_{\kappa m}(\bk') a^*_{\eta n}(\bp')\nonumber\\
&& \times \ \delta^{(2)}(\bk+\bp-\bp_X) \delta^{(2)}(\bk'+\bp'-\bp_X) {\cal M}(\bk,\bp) {\cal M}^*(\bk',\bp')\,.
\nonumber
\eea
\begin{figure*}
  \includegraphics[width=0.75\textwidth]{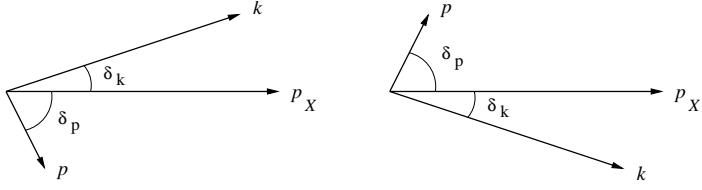}
\caption{Two kinematical configurations of the transverse momenta $\bk$ and $\bp$ of fixed absolute values
that sum up to the vector $\bp_X$.}
\label{fig1} 
\end{figure*}
Trying to satisfy all the kinematical restrictions which enter this expression at fixed final transverse momentum $\bp_X$,
we end up with exactly two kinematical configurations shown in Fig.~\ref{fig1}. These two configurations are at work 
both for $ {\cal M}(\bk,\bp)$ and the conjugate of ${\cal M}(\bk',\bp')$, which means that two relations
between $\bk, \bp$ and $\bk', \bp'$ are possible:
\bea
\mbox{direct:}&& \bk'=\bk\,,\ \bp'=\bp\,,\nonumber\\
\mbox{reflected:}&& \bk'=\bk^* \equiv - \bk + 2(\bk\bn_X)\bn_X \,,\ \bp'=\bp^* \equiv - \bp + 2(\bp\bn_X)\bn_X\,,\label{2configurations}
\eea
with $\bn_X \equiv \bp_X/|\bp_X|$.
Since these two possibilities interfere, the double-twisted cross section will depend not only
on $|{\cal M}(\bk,\bp)|^2$ but also on ${\cal M}(\bk,\bp) {\cal M}^*(\bk^*,\bp^*)$,
the {\em autocorrelation function} of the amplitude. Note that such a quantity is inaccessible with plane wave scattering.

Again, skipping the details which can be found in \cite{ivanov2011} we show the result for the cross section:
\be
d\sigma_{2tw} = {1 \over 8\pi \sin(\delta_k+\delta_p)} \int d\phi_k d\phi_p
{j_{PW}(\bk,\bp) \over j_{2tw}}\left[ d\sigma_{PW}(\bk,\bp)+ d\sigma'_{PW}(\bk,\bp)\right]\,,
\en
where $d\sigma_{PW}(\bk,\bp)$ is the usual plane wave cross section, while
\bea
d\sigma'_{PW}(\bk,\bp) &=& {(2\pi)^4\delta(E_i-E_f)\delta(p_{zi}-p_{zf})\delta^{(2)}(\bk+\bp-\bp_X)
\over 4 E_p \omega j_{PW}}\nonumber\\
&\times & \mathrm{Re}\left[e^{2im(\phi_k-\phi_X)+2in(\phi_p-\phi_X)}
{\cal M}(\bk,\bp){\cal M}^*(\bk^*,\bp^*)\right] d\Gamma_X\,.
\eea
and
$$
\delta_k = \arccos\left({\bp_X^2 + \kappa^2 -\eta^2 \over 2|\bp_X|\kappa}\right)\,,\quad 
\delta_p = \arccos\left({\bp_X^2 - \kappa^2 +\eta^2 \over 2|\bp_X|\eta}\right)\,.
$$
Note that, in contrast to the single-twisted case, the double-twisted cross section is $m,n$-dependent 
and, similarly to the single-twisted case, stays finite at small $\kappa$, $\eta$.

\section{Conclusions}

Orbital angular momentum (OAM) is a new degree of freedom, 
which can be used in high-energy physics to gain more insight into properties
of particles and their interactions.
In this contribution we discussed general aspects of high-energy collisions involving particles
with non-zero OAM in the initial state.

The single-twisted (i.e. OAM state collision with a plane wave) cross section is represented via the 
plane wave cross section averaged over the azimuthal angle of one of the incoming particles. 
If the initial twisted particle is prepared in a superposition state with different
orbital quantum numbers, a Fourier analysis of the plane wave cross section with respect to the initial azimuthal angle 
can be performed.
The expression we found for the double-twisted cross section involves not only 
the plane wave cross section, but also the autocorrelation function of the amplitude.
It is interesting now to see what new insights into specific high-energy processes can be inferred from these possibilities.

\begin{acknowledgements}
This work was supported by the Belgian Fund F.R.S.-FNRS via the contract of
Charg\'e de recherches, and in part by grants of the Russian
Foundation for Basic Research 09-02-00263-a and 11-02-00242-a
as well as NSh-3810.2010.2.\end{acknowledgements}

\end{document}